\documentclass[12pt,transaction, draftclsnofoot,onecolumn]{IEEEtran}
\usepackage{cite}
\usepackage{mathrsfs}
\usepackage{amsfonts,amssymb}
\usepackage{bm}
\usepackage{amsmath}
\usepackage{amssymb}
\usepackage{breqn}

\usepackage[ruled,linesnumbered]{algorithm2e}
\usepackage{graphicx}
\usepackage{subfigure}
\usepackage{epstopdf}
\usepackage{bbm}
\usepackage{amsthm}

\newtheorem{theorem}{Theorem}

\renewcommand\IEEEkeywordsname{Keywords}

\begin{document}
	
\title{Energy Efficient Mobile Edge Computing \\in Dense Cellular Networks}
%\author[1]{Lixing Chen}
%\author[2]{Sheng ZHou}
%\author[1]{Jie Xu}
%\affil[1]{Department of Electrical and Computer Engineering,
%\affil[2]{Department of Electronic Engineering, Tsinghua University, Beijing, China}

\author{\IEEEauthorblockN{Lixing Chen$^*$, Sheng Zhou$^\dagger$, Jie Xu$^*$}
\IEEEauthorblockA{$^*$Department of Electrical and Computer Engineering, University of Miami, USA\\
$^\dagger$Department of Electronic Engineering, Tsinghua University, China}}

\maketitle
\begin{abstract}
Merging Mobile Edge Computing (MEC), which is an emerging paradigm to meet the increasing computation demands from mobile devices, with the dense deployment of Base Stations (BSs), is foreseen as a key step towards the next generation mobile networks. However, new challenges arise for designing energy efficient networks since radio access resources and computing resources of BSs have to be jointly managed, and yet they are complexly coupled with traffic in both spatial and temporal domains. In this paper, we address the challenge of incorporating MEC into dense cellular networks, and propose an efficient online algorithm, called ENGINE (ENErgy constrained offloadINg and slEeping) which makes joint computation offloading and BS sleeping decisions in order to maximize the quality of service while keeping the energy consumption low. Our algorithm leverages Lyapunov optimization technique, works online and achieves a close-to-optimal performance without using future information. Our simulation results show that our algorithm can effectively reduce energy consumption without sacrificing the user quality of service.
\end{abstract}
%\begin{IEEEkeywords}
%	Mobile edge computing, multi-cell, computation offloading, base station sleeping
%\end{IEEEkeywords}
% \fi
%\IEEEpeerreviewmaketitle

\section{Introduction}
Many emerging mobile applications, such as mobile gaming and augmented reality, are delay sensitive and have resulted in an increasingly high computing demand that frequently exceeds what mobile devices can deliver. Although cloud computing enables convenient access to a centralized pool of configurable computing resources, moving all the distributed data and computing-intensive applications to clouds (which are often physically located in remote mega-scale data centers) is simply out of the question, since it would not only pose an extremely heavy burden on today's already-congested backbone networks but also result in (sometimes intolerable) large transmission latencies that degrade the quality of service. Mobile edge computing (MEC) (a.k.a. fog computing) thus has recently emerged as a remedy to the above limitations, which enables processing of (some) workloads locally at the network edge without moving them to the cloud \cite{beck2014mobile} \cite{vaquero2014finding}. In MEC, network edge devices, such as base stations (BSs), access points and routers, are endowed with, albeit limited, computing and storage capabilities to serve users' requests as a substitute of clouds, while significantly reducing the transmission latency as they are placed in the proximity of end users.

Although MEC promises enormous benefits, designing energy efficient (green) cellular networks faces significant new challenges. To accommodate the continuously growing demand for ubiquitous information access, BSs are becoming increasingly densely deployed. As a result, the energy consumption of BSs becomes a major portion (60\% - 80\%) of the whole cellular network energy consumption \cite{marsan2009optimal}, which is already one of the leading sources of the global carbon dioxide emissions. As one of the most popular and efficient energy saving schemes, BS sleeping has been proposed and widely studied to realize substantial energy saving in cellular networks \cite{bousia2012green}\cite{wu2013traffic}\cite{zhou2009green}. However, integrating MEC with BSs significantly complicates the energy saving issue due to the fact that BSs now provide not only radio access services but also computing services. First, since computing resources on BSs are limited, offloading some workload to the remote cloud is inevitable. As a result, the workload offloading decisions and the sleeping decisions have to be jointly considered for each BS. Second, the long-term energy consumption couples the offloading and BS sleeping decisions over time, and yet the decisions have to be made without foreseeing the future system dynamics (workload, wireless channel conditions etc.). Third, dense cellular networks create a complex multi-cell environment where the workload demand, radio resources and computing resources are highly coupled in both the spatial and the temporal domains. Effective resource management requires careful coordination among all BSs in the network, and decentralized solutions are much favored in order to reduce complexity.

\begin{figure}
	\centering
	\includegraphics[width=0.45\textwidth]{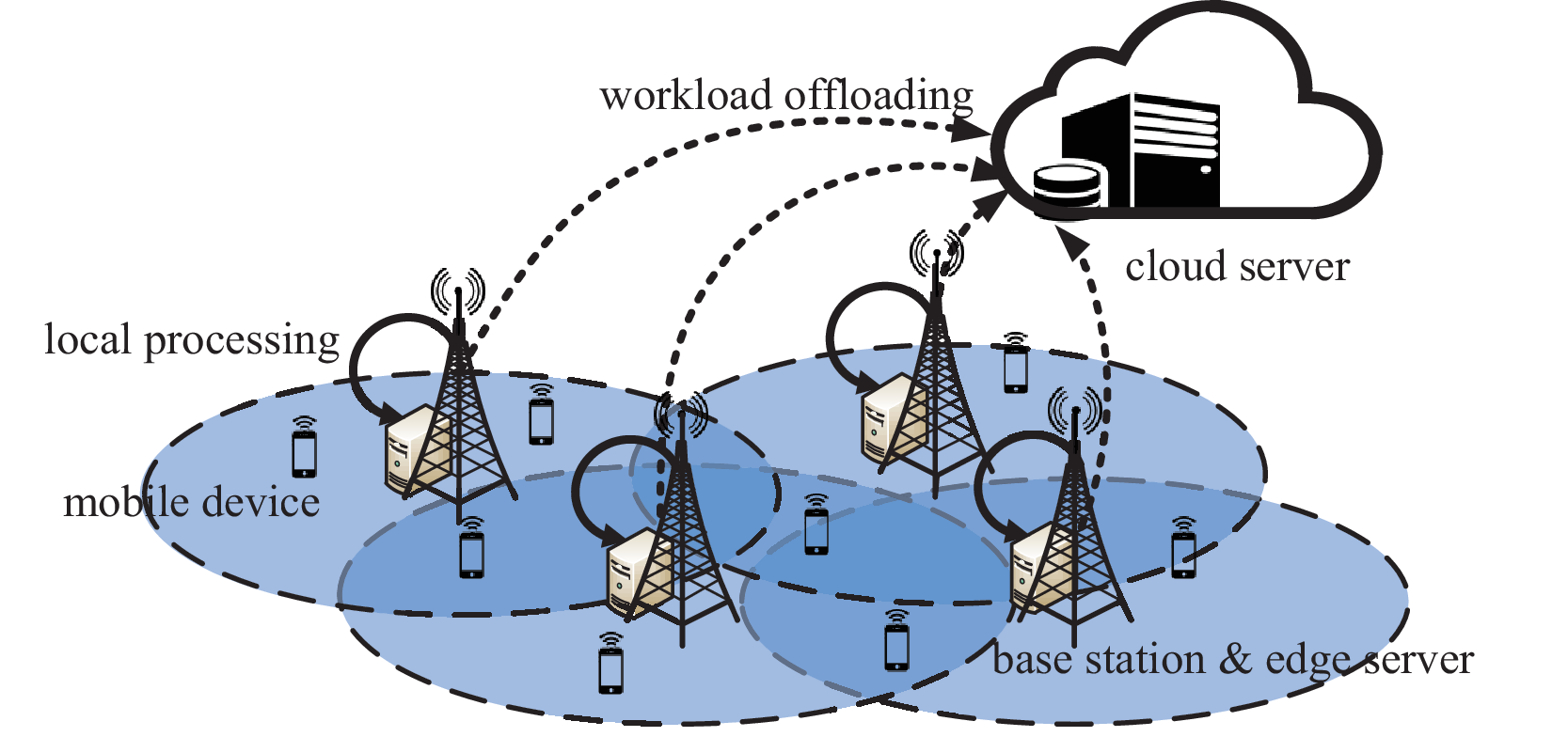}
	\caption{Scenario of multi-cell MEC}
	\label{scene}
\vspace{-0.2in}
\end{figure}

In this paper, we study the joint management of radio resources and computing resources in dense cellular networks with MEC integration in order to maximize the quality of service for users while keeping the energy consumption of the BSs low. Figure \ref{scene} illustrates the considered system.  Our main contributions are as follows:
\begin{itemize}
\item We formalize the joint workload offloading and BS sleeping problem in dense cellular networks with MEC integration, for maximizing the user quality of service under a long-term energy consumption constraint without foreseeing the future information. To our best knowledge, this is the first work that studies MEC offloading and BS sleeping in a coupled multi-cell network.
\item To solve this problem, we develop a novel online algorithm, called ENGINE (ENerGy efficient offloadINg and slEeping), by leveraging the Lyapunov optimization technique. We prove that our algorithm achieves a close-to-minimum delay cost to end users compared to the optimal algorithm with full future information, while bounding the potential violation of energy consumption constraint.
\item We develop a decentralized algorithm, called REJO (Random Evolving Joint Optimization), which is a key subroutine of ENGINE that enables efficient coordination among the BSs to optimize their sleeping and offloading decisions. This makes our algorithm scalable to large networks.
\end{itemize}

The rest of the paper is organized as follows. Section \ref{sec2} reviews the related work. In Section \ref{sec3}, we introduce the system model. Section \ref{sec4} formulates objects and constraints. Section \ref{sec5} focuses on online BS activation and traffic offloading with proposed algorithmic framework. Simulation and results are presented in Section \ref{sec6}. Conclusion is given in Section \ref{conclusion}.

\section{Related Work}\label{sec2}
Mobile edge computing has received an increasing amount of attentions in recent years. In particular, a central theme of many prior studies is offloading policy on the user side, i.e. what/when/how to offload a user's workload from its device to the edge system or cloud (see \cite{huang2012dynamic}\cite{satyanarayanan2009case} and references therein). Our work focuses on the edge-side offloading, which is much less studied in the literature, and hence complements these studies on user-side offloading. Our prior work \cite{xu2016online} studies joint edge-side offloading and autoscaling in renewable-powered MEC. However, the optimization for only one MEC device (BS) is considered. Offloading is much more difficult in a dense multi-cell environment since workload demands are complexly coupled among multiple BSs.

BS sleeping has been studied to realize substantial energy saving in green cellular networks since even a small reduction in the BS transmit power enables considerable savings in overall energy consumption due to its influence on the operational power of amplifiers, cooling systems etc. In some classical literature \cite{yadin1963queueing}\cite{heyman1968optimal}, ideas similar to the user number or vacation based sleeping design have been studied, where single server queueing analysis is carried out. Adopting the Markov decision process (MDP), authors of \cite{kamitsos2010optimal} prove that the optimal sleeping pattern of serving delay-tolerant jobs for a typical server has a simple hysteretic structure. Our prior work \cite{wu2013traffic}\cite{wu2016base} designs joint sleeping and power matching schemes for energy-delay tradeoffs with non-realtime traffic arrival at a single BS. For the multi-cell scenario, \cite{zhou2009green} designs BS sleeping schemes in dense cellular networks considering the randomness and the spatial distribution of traffic. However, computation offloading for MEC is barely considered in existing works.

Jointly optimizing radio and computational resources for multicell MEC is studied in \cite{sardellitti2015joint}. In this work, however, although radio resources are distributed among multiple cells, all computation workload is processed at a single cloud server. This is significantly different from our setting in which computation workload is also processed locally at each BS whenever possible.

\section{System Model}\label{sec3}
We consider a densely-deployed wireless system with $N$ BSs, indexed by $\mathcal{N}\in\{1,2,\ldots,N\}$. The operational time line is divided into discrete time slots. As a major deployment method of MEC, we consider that each BS is co-located with an edge server, and shares the same power supply with it.

\subsection{Traffic Model}
The network is divided into $M$ disjoint regions, indexed by $\mathcal{M}\in\{1,2,\ldots,M\}$, In time slot $t$, the amount of traffic arrival to region $m$ is denoted by $\lambda^t_m\in\mathbb{R}_+$. Among this traffic, $\rho\in(0,1)$ fraction is computation traffic and the rest is pure communication traffic. In this paper, we assume $\rho$ is homogeneous over the whole network and across time. $\bm{\lambda}^t=(\lambda^t_1,...,\lambda^t_M)$ denotes the overall traffic pattern across all regions.
Each region $m$ can be covered by a set of BSs, denoted by $\mathcal{B}_m\subseteq\mathcal{N}$, due to the dense deployment of BSs. BSs can be in either the active mode or the sleeping mode. Let $a^t_n\in\{0,1\}$ represent the active (1)/sleeping (0) decision for BS $n$ in time slot $t$. Let $\mathcal{A}^t_m\subseteq\mathcal{B}_m$ denote the set of active BSs serving region $m$ in time slot $t$. For analytical simplicity, we assume that the traffic $\lambda^t_m$ in region $m$ is equally distributed among active BSs. Load balancing among the active BSs is our future work. Therefore, the traffic arrival $\mu^t_n(\bm{a}^t)$ to a BS $n$ is
%\begin{equation}
%\mu^t_n(a^t) = \left\{
%        \begin{array}{lcl}
%           {\sum\limits_{m=1}^{M}\mathbbm{1}\{n\in\mathcal{B}_m\}\dfrac{\lambda^t_m}{|\mathcal{A}^t_m|}} &,\text{if} & a^t_n=1 \\
%           \quad {0} &,\text{if} & a^t_n=0
%           \end{array}\right.
%\end{equation}

\begin{equation}
\mu^t_n(\bm{a}^t) =      a^t_n\sum\limits_{m=1}^{M}\textbf{1}\{n\in\mathcal{B}_m\}\frac{\lambda^t_m}{|\mathcal{A}^t_m|}
\end{equation}
The computation traffic can be processed at the local edge server or offloaded to a remote cloud. For an activated BS $n$, let $b^t_n\in[0,1]$ denote the fraction of computation workload processed at edge server, which will impact the power consumption and delay cost of MEC systems as we will model next.

\subsection{Power Consumption Model}
The power consumption consists of \emph{operational power} $P^t_{op}$, \emph{transmission power} $P^t_{tx}$ and \emph{computation power} $P^t_{com}$.

The operational power is load-independent, consisting of the baseband processor, the converter, the cooling system and etc. When BS $n$ is in the sleeping mode, the operational power becomes $0$. Therefore, for BS $n$ in time slot $t$:
\begin{equation}
P^t_{op,n}=P_0a^t_n,~P_0~\text{is a constant.}
\end{equation}	

Transmission occurs on both the wireless link between end users and BSs, and the wired link between BSs and remote cloud. Usually the wireless transmission power consumption dominates and hence we consider only the wireless part. Since the considered time slot is relatively long, we assume that small-scale fast fading will average out. Hence, we focus on pathloss effects. By making each region small, we can approximate the pathloss effect by considering the average distance between BS $n$ and region $m$, denoted by $d_{n,m}$. Given transmission power $P_{n,m}$, the maximum achievable transmission rate is given by the Shannon channel capacity,
\begin{equation}
r_{n,m}=W\log_2\left(1+\frac{P_{n,m}\beta(d_{n,m}^{-\alpha})}{\sigma^2}\right)
\end{equation}
where $W$ is the channel bandwidth, $\beta$ is the pathloss constant, $\alpha$ is the pathloss exponent, and $\sigma^2$ is the noise power. We consider the noise-limited setting by assuming that BSs operate on orthogonal channels. Suppose each transmission must meet a target rate $r_0$ to satisfy a transmission delay requirement, then the transmission power must satisfy:
\begin{equation}
P_{n,m}=(2^{\frac{r_0}{W}}-1)\sigma^2(d_{n,m})^\alpha\beta^{-1}
\end{equation}
The transmission power of BS $n$ in time slot $t$ is thus
\begin{equation}
P^t_{tx,n}=a^t_n\sum\limits_{m:n\in\mathcal{B}_m}^{}\frac{\lambda^t_m}{\left|\mathcal{A}^t_m\right|}(2^{\frac{r_0}{W}}-1)\sigma^2(d_{n,m})^\alpha\beta^{-1}
\end{equation}
The computation power at edge server is load-dependent. Let
\begin{equation}
 P^t_{com,n}=g\left(\rho b^t_n\mu^t_n(\bm{a}^t)\right)
\end{equation}
 denote the computation power of BS $n$ to process local computation workload $\rho b^t_n\mu^t_n(\bm{a}^t)$, where $g(\bm{\cdotp})$ is assumed to be an increasing function.

\subsection{Delay Cost Model}
For local processed workload, the delay cost $c^t_{lo,n}$ is mainly the processing delay due to the limited computing capacity at edge servers. The transmission delay from BS to edge server is negligible due to physical colocation. To quantify the delay performance of services without restricting our model to any particular metric, we use a general notion to represent $c^t_{lo,n}$, modeling the service process as a M/M/1/PS queue and using average response time to represent the delay cost \cite{ren2013coca}:
\begin{equation}
c^t_{lo,n}=\frac{\rho b^t_n\mu^t_n(\bm{a}^t)}{\chi_n-\rho b^t_n\mu^t_n(\bm{a}^t)}
\end{equation}
where $\chi_n$ is the maximal service rate of BS $n$.

For offloaded computation workload, the delay cost $c^t_{rem},n$ is mainly transmission delay due to network round trip time (RTT), which depends on the network congestion state. For modeling simplicity, the service time at the cloud side is also absorbed into the network congestion state. Thus we model the network congestion state $h^t_n$ as an exogenous parameter and express it in terms of RTT. Therefore,
\begin{equation}
c^t_{rem,n}=\rho (1-b^t_n)\mu^t_n(\bm{a}^t)h^t_n
\end{equation}

\section{Problem Formulation}\label{sec4}
\subsection{Objective and Constraints}
The network operator considers the traffic arrival pattern $\bm{\lambda}^t$ and the network congestion state $\bm{h}^t$ as inputs, and decides BS activation strategy $\bm{a}^t$ and offloading strategy $\bm{b}^t$. It aims at maximizing the QoE subject to a set of constraints, as specified below.

\textbf{Objective}: Since MEC is mainly concerned with the delay performance, the optimization objective is formulated to minimize the average delay cost expressed as:
\begin{equation}
\bar{c}=\frac{1}{T}\sum\limits_{t=1}^{T}\sum\limits_{n=1}^{N}(c^t_{lo,n}(\bm{a}^t,\bm{b}^t,\bm{\lambda}^t)+c^t_{rem,}(\bm{a}^t,\bm{b}^t,\bm{\lambda}^t,\bm{h}^t))
\end{equation}

\textbf{Constraints}: To cover the whole network, BS activation decisions need to satisfy
\begin{equation}
\sum\limits_{n\in\mathcal{B}_m}^{}a^t_n\geq 1,~\forall m,~\forall t
\end{equation}

To avoid severe offloading and workload dropping, the offloading decisions need to satisfy
\begin{equation}
b^t_n\mu^t_n(\bm{a}^t)\leq \gamma\bm{\cdot}\chi_n,~\forall n,~\forall t
\end{equation}
where $\gamma\in(0,1)$ is a predetermined parameter that controls the maximum utilization of edge servers.

The per-time slot power consumption of each BS $n$ is capped by an upper limit $\bar{P}_n$
\begin{equation}
P^t_{op,n}(a^t_n)+P^t_{tx,n}(\bm{a}^t,\bm{\lambda}^t)+P^t_{com,n}(\bm{a}^t,\bm{b}^t,\bm{\lambda}^t)\leq\bar{P}_n
\end{equation}

The network operator has a long-term energy consumption budget. Mathematically, the network operator desires to follow the long-term constraint specified by
\begin{equation}
\begin{split}	
\frac{1}{T}\sum\limits_{t=1}^{T}\sum\limits_{n=1}^{N}(P^t_{op,n}&(a^t_n)+P^t_{tx,n}(\bm{a}^t,\bm{\lambda}^t)\\
&+P^t_{com,n}(\bm{a}^t,\bm{b}^t,\bm{\lambda}^t))\leq Q
\end{split}
\end{equation}
\subsection{Offline Problem Formulation}
The offline problem $\bm{\mathcal{P}1}$ is formulated as follows:
\begin{equation*}
\begin{split}
\min\limits_{\bm{a},\bm{b}}\bar{c}=&\frac{1}{T}\sum\limits_{t=1}^{T}\sum\limits_{n=1}^{N}(c^t_{lo,n}(\bm{a}^t,\bm{b}^t,\bm{\lambda}^t)+c^t_{rem,n}(\bm{a}^t,\bm{b}^t,\bm{\lambda}^t,\bm{h}^t))\\
&s.t.\qquad \text{constraints}\quad(10),(11),(12),(13) \\
\end{split}
\end{equation*}
Optimally solving $\bm{\mathcal{P}1}$ requires complete offline information which are difficult to predict in advance, if not impossible. Moreover, $\bm{\mathcal{P}1}$ is a mixed integer nonlinear programming and is very difficult to solve even if the future information is known a priori. These challenges demand an online approach that can efficiently achieve the joint optimization.

\section{Online BS Activation and Traffic Offloading}\label{sec5}
In this section, we develop online algorithms to jointly optimize BS activation strategy $\bm{a}^t$ and offloading $\bm{b}^t$ strategies.
\subsection{Lyapunov optimization based online algorithm}
Our algorithm, called ENGINE, solves $\bm{\mathcal{P}1}$ based on Lyapunov optimization technique \cite{neely2010stochastic}. The algorithm is purely online and requires only currently available information as inputs. Specifically, we construct a virtual power deficit queue $q(t)$ which guides the decision to follow the long-term power consumption constraint. The power deficit queue evolves as follows:
\begin{equation}
q(t+1)=\max\left(q(t)+P^t-Q,0\right),~q(0)=0
\end{equation}
where  $P^t=\sum^N_{n=1}(P^t_{op,n}+P^t_{tx,n}+P^t_{com,n})$, the length of $q(t)$ indicates the deviation of current power consumption from the power consumption constraint. ENGINE is presented in algorithm 1, where $c^t= c^t_{lo,n}+c^t_{rem,n}$.
Theorem 1 provides the performance guarantee of ENGINE.

\begin{algorithm}[hbt]
	\caption{ENGINE}
	\KwIn{Constraint Values $\bar{P}_n$, $Q$}
	\KwOut{BS activation strategy $\bm{a}^t$, offloading scheme $\bm{b}^t$}
	$q(0)\leftarrow0$\;
	\For{$t$=1 \emph{\textbf{to}} $T$}
	{
	Observe $\bm{\lambda}^t$,$\bm{h}^t$ at the beginning of each time slot t\;
	Choose  $\bm{a}^t$, $\bm{b}^t$ to minimize\\ \quad$\bm{\mathcal{P}2}$: $V\bm{\cdot}c^t(\bm{a}^t,\bm{b}^t,\bm{\lambda}^t,\bm{h}^t)+q(t)\bm{\cdot}P^t(\bm{a}^t,\bm{b}^t,\bm{\lambda}^t)$\;
		Update $q(t+1)\leftarrow \max\left(q(t)+P^t-Q,0\right)$\;
	}
	return $\bm{a}^t$,$~\bm{b}^t$\;
\end{algorithm}

\begin{theorem}
By applying ENGINE, the long-term average delay cost satisfies:
\begin{equation}
\lim\limits_{T\rightarrow\infty}\frac{1}{T}\sum\limits_{t=1}^{T}\mathbb{E} \{c^t\}\leq c^*+\frac{1}{2V}(\sum\limits_{n=1}^{N}\bar{P}_n-Q)^2
\end{equation}
and the long term average power consumption satisfies:
\begin{equation}
\lim\limits_{T\rightarrow\infty}\frac{1}{T}\sum\limits_{t=1}^{T}\mathbb{E} \{P^t\}\leq \frac{(\sum\limits_{n=1}^{N}\bar{P}_n-Q)^2+2V c^*}{2(Q-P^{*,t})}+Q
\end{equation}
\end{theorem}
\begin{proof}
See Appendix.
\end{proof}

Theorem 1 proves a strong performance guarantee for ENGINE: the long-term delay cost is upper-bounded by the optimal delay cost $c^*$ plus a constant. The long term power consumption is no larger than the constraint $Q$ plus a constant. Both constants depend on the control parameter $V$, which makes trade-off between delay cost and power consumption.
\subsection{BS activation and traffic offloading joint optimization}
In this part, we focus on solving $\bm{\mathcal{P}2}$ to find the optimal BS activation $\bm{a}^t$ and offloading $\bm{b}^t$ strategies for each time slot $t$. $\bm{\mathcal{P}2}$ is a joint optimization problem which can be solved via centralized techniques using greedy searching. However these methods are usually computationally prohibitive, and in practice distributed solutions are preferred. We propose an efficient algorithm that enables decentralized implementation, called Random Evolving Joint Optimization (REJO), based on the Gibbs Sampling technique \cite{robert2013monte}, which is presented in \emph{Algorithm} 2. For each time slot $t$, the optimal solution pair ($\bm{a}^{opt}, \bm{b}^{opt}$) is found in an iterative manner. In each iteration, a randomly chosen BS $n$ virtually evolves its working mode $a^t_n$. The corresponding optimal offloading scheme $b^t_n$ is derived by minimizing the objective function in Line 5 using the activation and offloading strategies of other BSs (i.e. $\bm{\tilde{a}}^t_{-n}$, $\bm{\tilde{b}}^t_{-n}$). The key feature of REJO is the randomness introduced in the decision making (Line 8 and 9). Specifically, BS $n$ may maintain its current mode or explore with a certain probability the other mode to avoid being trapped in local optimal solution. The parameter $\tau$ is used to control the probability of exploring. When $\tau$ is small, REJO tends to keep the current solution and therefore may be stuck in a local optimal solution. When $\tau$ is large, REJO explores all possible solutions and therefore it takes more time to converge.

\begin{algorithm}[!h]
	\caption{REJO (time slot $t$)}
	\KwIn{ $\bm{\tilde{a}}\leftarrow\bm{a}^{t-1}$, $\bm{\tilde{b}}\leftarrow\bm{0}$, $o^{opt}\leftarrow o^{t-1}$, $q(t)$}
	\KwOut{BS activation strategy $\bm{a}^t$, offloading scheme $\bm{b}^t$}
	\While{stoping criterion is NOT satisfied}
	{
		Randomly pick BS $n$ and select working mode $a_n$\;
		$\bm{\tilde{a}}\leftarrow(\bm{\tilde{a}}_{-n},a_n)$\;
		\If{$\bm{\tilde{a}}$ is feasible}  {
			Choose $b_n$ by minimizing $V\bm{\cdot}c^t(\bm{\tilde{a}},\bm{\tilde{b}}_{-n},b_n,\bm{\lambda}^t,\bm{h}^t)+q(t)\bm{\cdot}P^t(\bm{\tilde{a}},\bm{\tilde{b}}_{-n},b_n,\bm{\lambda}^t)$\; $\bm{\tilde{b}}\leftarrow(\bm{\tilde{b}}_{-n},b_{n})$\;
			$\tilde{o}\leftarrow
			c^t(\bm{\tilde{a}},\bm{\tilde{b}},\bm{\lambda}^t,\bm{h}^t)+q(t)\bm{\cdot}P^t(\bm{\tilde{a}},\bm{\tilde{b}},\bm{\lambda}^t)$\;
			$k\leftarrow\frac{1}{1+e^{(\tilde{o}-o^{opt})/\tau}}$\;
			With probability of $k$, BS $n$ set $a^{opt}_n\leftarrow a_n$, $b^{opt}_n\leftarrow b_n$, $o^{opt}_n\leftarrow \tilde{o}$, broadcast $a^{opt}_n$, $b^{opt}_n$, $o^{opt}_n$
			\;
		}
	}
	$\bm{a}^t\leftarrow\bm{a}^{opt}$, $\bm{b}^t\leftarrow\bm{b}^{opt}$\;
\end{algorithm}

An important feature of REJO is that it enables decentralized implementation which allows each BS to make autonomous decisions. Since in each iteration only one BS is chosen to evolve its working mode, the chosen BS is able to optimize its offloading scheme locally. After each iteration, the chosen BS $n$ communicates $a^t_n$ and $b^t_n$ to other BSs, which prepares for subsequent iterations. Next, we formally prove the feasibility of our algorithm.

\begin{theorem}
As $\tau$ decreases, REJO converges with a higher probability to the global optimal solution. When $\tau\rightarrow0$, REJO converges to the global optimal solution with probability of 1.
\end{theorem}
\begin{proof}
See Appendix.
\end{proof}

\section{Simulation}\label{sec6}
\subsection{Simulation Setup}
\begin{figure}
\centering
\includegraphics[width=2 in]{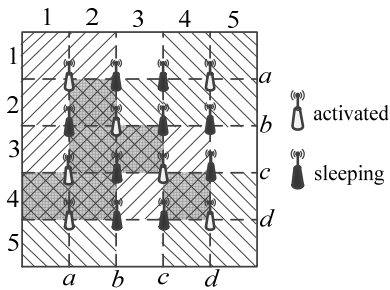}
\vspace{-0.2in}
\caption{Simulation model}
\label{SimuMo}
\vspace{-0.2in}
\end{figure}

The simulation model is given in Fig. \ref{SimuMo}. We consider a $5\times5$ grid of square regions covered by 16 BSs located on the grid intersection. The BSs are densely-deployed and hence they have overlapping coverage areas. The coverage radius of each BS is 1. Each region must be covered by at least one activated BS and we assume that the traffic arrival rate in each region is normally distributed and traffic load is equally distributed among the activated BSs $\mathcal{A}^t_m$ severing the region.

\subsection{Results}
\subsubsection{Performance and Comparison}

\begin{figure}
\centering
	\subfigure[Average delay cost]{\label{aveDelay}
		\includegraphics[width=1.6in]{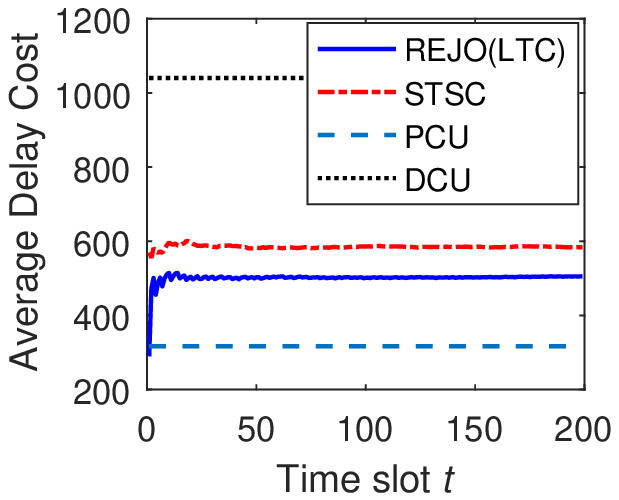}}
	  \hspace{0in}
	\subfigure[Average power consumption]{\label{avePow}
		\includegraphics[width=1.6in]{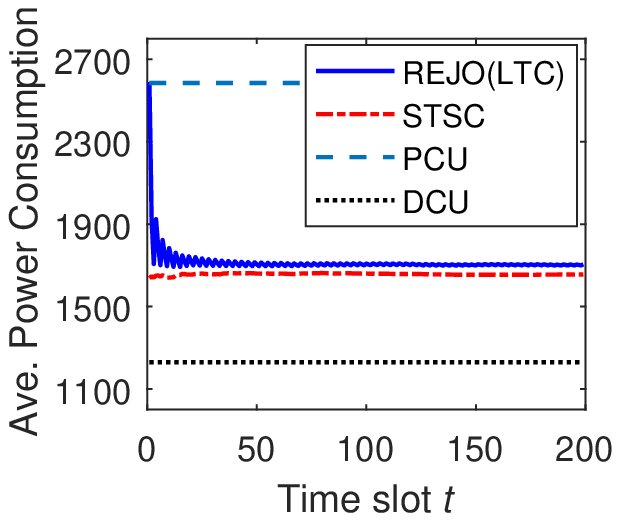}}
	%\hspace{\fill}
	\caption{Performance of REJO ($V=200$, $Q=1750)$}
	\label{aveP}
\vspace{-0.1in}
\end{figure}
\begin{figure}
	\centering
\vspace{-0.1in}
	\includegraphics[width=3in]{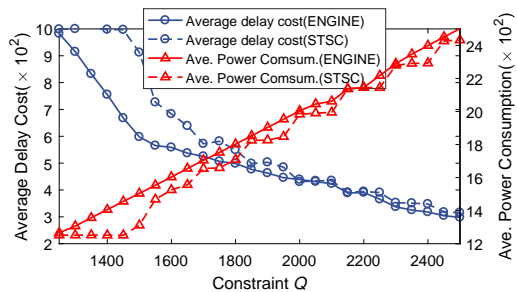}
\vspace{-0.1in}
	\caption{Impact of $Q$}
	\label{varyingQ}
\vspace{-0.2in}
\end{figure}

Fig. \ref{aveP} illustrates the average power consumption $\bar{P}$ and average delay cost $\bar{c}$ across 200 time slots. We compare ENGINE with three benchmark methods: Single Time Slot Constraint (STSC), Power Consumption Unaware (PCU) and Delay Cost Unaware (DCU). STSC exerts power consumption constraint on each time slot instead of using Long Term Constraint (LTC). Compared to STSC, ENGINE achieves smaller $\bar{c}$ under similar $\bar{P}$. PCU incurs smaller delay cost $\bar{c}$ yet causes larger power consumption since it optimizes delay cost while ignoring the power consumption. DCU focuses on power saving and disregards the delay cost. It achieves a lower $\bar{P}$ at a much higher delay cost $\bar{c}$. In Fig. \ref{varyingQ}, we illustrate the trade-off between $\bar{c}$ and $\bar{P}$ for various power consumption constraints $Q$. It can be seen that when the energy constraint $Q$ is loosened, ENGINE adapts itself and achieves a lower delay cost $\bar{c}$.

\subsubsection{BS activation and traffic offloading}
\begin{figure}
	\centering	
	\subfigure[Offloading scheme]{\label{offloading}
		\includegraphics[width=1.6in]{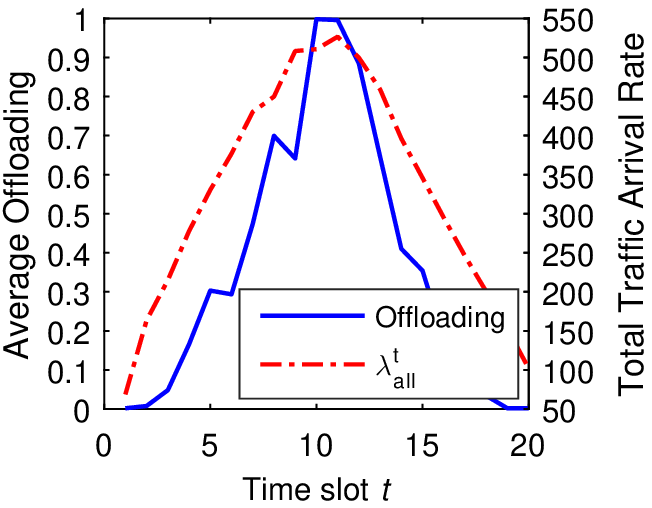}}	
	\subfigure[BS activation]{\label{lambdaActi}
		\includegraphics[width=1.5in]{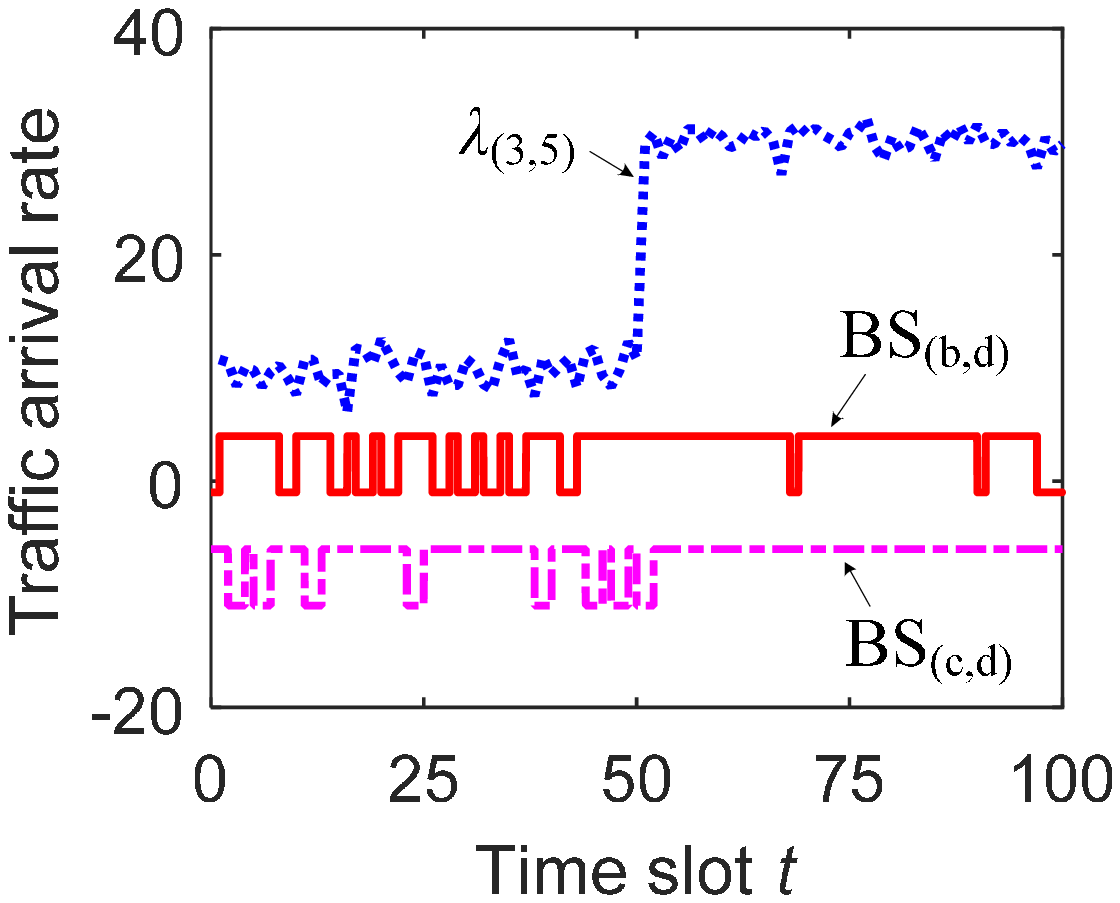}}
\vspace{-0.1in}
	\caption{Influence of traffic arrival rate}
	\label{traffic}
\end{figure}

Fig. \ref{traffic} shows the impact of the traffic arrival rate on offloading and BS activation decisions. In Fig. \ref{offloading}, we gradually increase the traffic arrival rate in each region, and observe the average offloading decisions of BSs. The result is intuitive: when the traffic arrival rate is high, BS tends to offload more computation load in order to avoid high computation delay and power consumption at edge servers.

Fig. \ref{lambdaActi} illustrates the relation between the traffic arrival rate and the BS activation decision. For better presentation we only alter traffic arrival rate $\lambda_{(3,5)}$ in Region(3,5) and observe the working mode of BS$_{\text{(b,d)}}$ and BS$_{\text{(c,d)}}$. As the result shows, when $\lambda_{(3,5)}$ is small, only one BS tends to be activated; after the increase of $\lambda_{(3,5)}$ in 50-th time slot, both BSs are activated in almost all time slots.

\begin{figure}
	\begin{minipage}[t]{0.5\linewidth}
		\centering
		\includegraphics[width=1.5in]{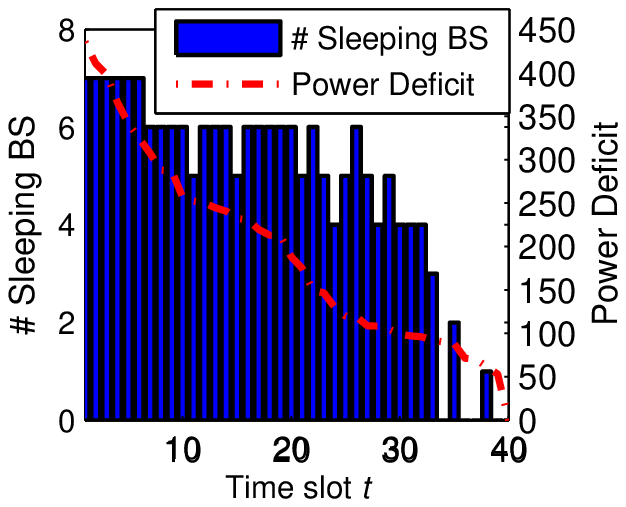}
\vspace{-0.2in}
		\caption{Impact of $q$ on BS sleeping}
		\label{ActiDeficit}
	\end{minipage}%
	\begin{minipage}[t]{0.5\linewidth}
		\centering
		\includegraphics[width=1.5in]{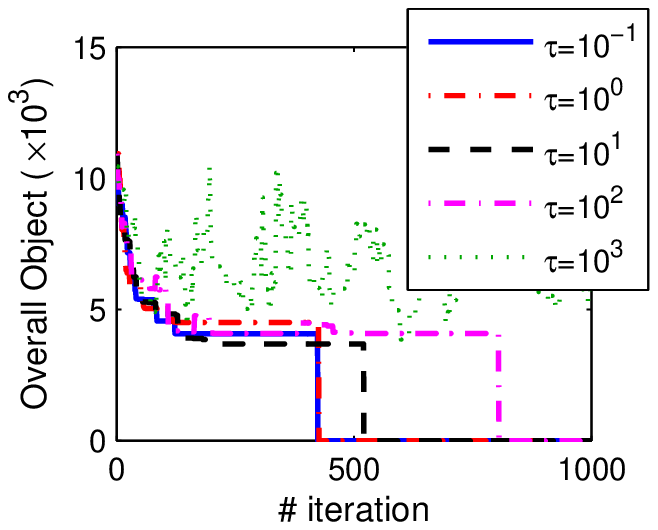}
\vspace{-0.2in}
		\caption{Execution of random evolving}
		\label{tau}
	\end{minipage}
\end{figure}

Fig. \ref{ActiDeficit} shows the BS activation decisions under different power deficit $q$. The result indicates that when $q$ is large, more BSs are switched off to cut the power deficit. When $q$ is small, more BSs are actived to minimize delay cost. Notice that the maximum number of sleeping BSs is 7, since at least 9 BSs must be activated to cover all the regions.
\subsubsection{Execution of Random Evolving}
Fig. \ref{tau} shows the evolution of objective values ($\bm{\mathcal{P}}_2$) during iterations. The result matches our analysis: with a smaller $\tau$, the evolution converges fast while being potentially trapped in a local optimal solution; As $\tau$ grows, the evolution takes more time to converge or even does not converge, e.g. $\tau=10^3$.
\section{Conclusions}\label{conclusion}
In this paper, we focused on MEC in a multi-cell network and jointly optimized offloading and BS sleeping strategies for minimizing the computation delay cost while satisfying a power consumption constraint. Efficient online algorithms ENGINE and REJO were proposed to solve this problem without knowing the future information. Future works include load balancing among multiple active BSs and studying the dual problem of minimizing power consumption given the computation delay constraint.

%\section*{Acknowledgment}

\appendix
\textbf{Proof of Theorem 1}. Define the Lyapunov function $L\left(q(t)\right)\triangleq\frac{1}{2}q^2(t)$. Let the drift be $\Delta\left(q(t)\right)\triangleq L\left(q(t+1)\right)-L\left(q(t)\right)$. We have
\begin{equation}\label{drift}
\begin{split}
\Delta\left(q(t)\right)&=\frac{1}{2}\left(q^2(t+1)-q^2(t)\right)\\
&\leq\frac{1}{2}\left[\left(q(t)+P^t-Q\right)^2-q^2(t)\right]\\
&=\frac{1}{2}(P^t-Q)^2+(P^t-Q)q(t)
\end{split}
\end{equation}
%The inequality in (\ref{drift}) comes from $(\left[q(t)+P^t-Q\right]^+)^2\leq \left(q(t)+P^t-Q\right)^2$.
Let $\bm{a}^*$, $\bm{b}^*$ denote the BS activation strategy and traffic offloading scheme that generate the lowest delay cost $c^*$. let $\bm{a}^t$, $\bm{b}^t$ denote the optimal BS activation and offloading scheme minimizing $\bm{\mathcal{P}2}$. We have the \emph{drift-plus-penalty} expression:
\begin{equation}
\begin{split}
\Delta(q(t)&)+V c^t\leq\frac{1}{2}(P^t-Q)^2+V c^t+(P^t-Q)q(t)\\
&\leq\frac{1}{2}(\sum\limits_{n=1}^{N}\bar{P}_n-Q)^2+V c^t+(P^t-Q)q(t)\\
&\leq\frac{1}{2}(\sum\limits_{n=1}^{N}\bar{P}_n-Q)^2+V c^*+(P^{*,t}-Q)q(t)\\
\end{split}
\end{equation}
This is in the exact form for application of the \emph{Lyapunov Optimization Theorem} \cite{neely2010stochastic}, and therefore we obtain the claimed results. Notice that when $T\rightarrow\infty$, $\frac{1}{T}\sum_{t=1}^{T}\mathbb{E}\{q(t)\}\geq \frac{1}{T}\sum_{t=1}^{T}\mathbb{E}\{P^t\}-Q$, which gives upper-bound of long term power consumption.

\textbf{Proof of Theorem 2}.
For notational convenience, we drop the time index. Following the iterations in REJO, $\bm{a}$ evolves as a $N$-dimension Markov Chain. We first use 2-BS case, let $S_{i,j}$ denote the state of $\{a_1=i, a_2=j\}$, $i,j\in\{0,1\}$. Since each iteration only one BS is allowed to evolve, we have
\begin{equation}\label{PR1}
\begin{split}
\text{Pr}&(S_{m,n}|S_{i,j})=\\
&\left\{
\begin{array}{lcl}
{\frac{e^{-o\left(S_{m,n}\right)/\tau}}{2|a|(e^{-o\left(S_{m,n}\right)/\tau}+e^{-o\left(S_{i,j}\right)/\tau})}}, & m=i\text{ or }n=j \\
{0}, &\text{otherwise}
\end{array}\right.
\end{split}
\end{equation}
where $|a|=2$ is the size of BS action set and $o(S_{i,j})$ is the object value of BS state $S_{i,j}$. Then we have the balanced equations:
\begin{align}\label{PR2}
\begin{split}
\text{Pr}^*\left(S_{1,1}\right)&\text{Pr}^*\left(S_{1,2}|S_{1,1}\right)=
\text{Pr}^*\left(S_{1,2}\right)\text{Pr}^*\left(S_{1,1}|S_{1,k}\right)
\end{split}
\end{align}
Combining (\ref{PR1}) and (\ref{PR2}), we have
\begin{equation}\label{PR3}
\begin{split}
\text{Pr}^*&\left(S_{1,1}\right)\times\frac{e^{-o\left(S_{1,2}\right)/\tau}}{\left(e^{-o\left(S_{1,1}\right)/\tau}+e^{-o\left(S_{1,2}\right)/\tau}\right)}=\\
&\text{Pr}^*\left(S_{1,2}\right)\times\frac{e^{-o\left(S_{1,1}\right)/\tau}}{\left(e^{-o\left(S_{1,1}\right)/\tau}+e^{-o\left(S_{1,2}\right)/\tau}\right)}
\end{split}
\end{equation}

Observing the symmetry of equation (\ref{PR3}) as well as the Markovian chain, we note it can be applied for arbitrary state $\tilde{S}$ in the strategy space $\Omega$, and the stationary distribution is: $\text{Pr}^*(\tilde{S})=Ke^{-o\left(\tilde{S}\right)/\tau}$, where $K$ is a constant. Applying the probability conservation law, we have the stationary distribution for Markovian chain:
\begin{equation}\label{PR4}
\text{Pr}^*(\tilde{S})=\frac{e^{-o\left(\tilde{S}\right)/\tau}}{\sum_{S_i\in\Omega}e^{-o\left(\tilde{S}_i\right)/\tau}}
\end{equation}
Let $S^*$ be the optimal state yielding minimum value of object function, i.e. $S^*=\arg\min_{S_i\in\Omega}o(S_i)$. From (\ref{PR4}), we have $\lim\limits_{\tau\rightarrow0}\text{Pr}^*(S^*)=1$. The analogous analysis can be straightforwardly extended to an $N$-dimensional Markovian chain, thus completes the proof.


%\bibliographystyle{IEEEtran}

\bibliographystyle{IEEEtran}
\bibliography{refs}

\end{document}